\documentclass[12pt]{article}
\usepackage{graphicx,a4,epsfig,here,rotate,amsmath}    
\usepackage{latexsym,color,citesort,amssymb}
\usepackage{booktabs}

\newcommand{\be}{\begin{equation}}
\newcommand{\ee}{\end{equation}}
\newcommand{\eq}{\begin{eqnarray}}
\newcommand{\en}{\end{eqnarray}}
\newcommand{\bea}{\begin{eqnarray}}
\newcommand{\eea}{\end{eqnarray}}

\newcommand{\ed}{\end{document}}
\newcommand{\bc}{\begin{center}}
\newcommand{\ec}{\end{center}}
\newcommand{\sMpi}{\!\stackrel{\mbox{\tiny\sf o}}{M}_\pi^2}
\newcommand{\sMK}{\!\stackrel{\mbox{\tiny\sf o}}{M}_K^2}
\newcommand{\sMeta}{\!\stackrel{\mbox{\tiny\sf o}}{M}_\eta^2}
\newcommand{\sMP}{\!\stackrel{\mbox{\tiny\sf o}}{M}_P^2}
\newcommand{\sMV}{\!\stackrel{\mbox{\tiny\sf o}}{M}_V^2}

\begin{document}

\thispagestyle{empty}
\begin{center}

\vspace{3cm}
{\Large{\bf Feynman-Hellmann theorem for resonances\\[0.3em] and the 
%nature of certain states in QCD
quest for  QCD exotica
}}

\vspace{0.5cm}
\today

\vspace{0.5cm}
J.~Ruiz~de~Elvira$^a$,
U.-G.~Mei{\ss}ner$^{b,c}$,
A.~Rusetsky$^b$ and
G.~Schierholz$^d$

\vspace{2em}

\begin{tabular}{c}
$^a\,${\it Albert Einstein Center for Fundamental Physics,}\\
{\it Institute for Theoretical Physics, University of Bern,}\\
{\it Sidlerstrasse 5, 3012 Bern, Switzerland}\\[2mm]
$^b\,${\it Helmholtz--Institut f\"ur Strahlen-- und Kernphysik and}\\
{\it Bethe Center for Theoretical Physics,}\\
{\it Universit\"at Bonn, D--53115 Bonn, Germany}\\[2mm]
$^c${\it Institute for Advanced Simulation (IAS-4), 
Institut f\"ur Kernphysik  (IKP-3),}\\
{\it J\"ulich Center for Hadron Physics and JARA-HPC}\\
{\it Forschungszentrum J\"ulich, D-52425 J\"ulich, Germany}\\[2mm]
$^d$ {\it Deutsches Elektronen-Synchrotron DESY,}\\
{\it D-22603 Hamburg, Germany}

\end{tabular} 

\end{center}

\vspace{1cm}

{\abstract
The generalization of the Feynman-Hellmann theorem for resonance states in
quantum field theory is derived. On the basis of this theorem, 
a criterion is proposed to study the possible exotic nature of certain hadronic states 
emerging in  QCD. It is shown that this proposal is supported by explicit calculations 
in Chiral  Perturbation Theory and by large-$N_c$ arguments. Analyzing  recent
lattice data on the quark mass dependence in the pseudoscalar, vector meson,
baryon octet and baryon decuplet sectors, we conclude that, as expected, 
these are predominately quark-model states, albeit the corrections are non-negligible.

}

\clearpage

%\tableofcontents

\section{Introduction}

The celebrated Feynman-Hellmann theorem~\cite{Hellmann,Feynman:1939zza}
addresses the situation when a quantum-mechanical Hamiltonian $H(\lambda)$
of a given system depends on a some external parameter $\lambda$. 
The energy spectrum $E_n(\lambda)$ and the wave functions 
$|\Psi_n(\lambda)\rangle$ will then depend on this parameter as well. The theorem relates
the $\lambda$-dependence of the energy spectrum to the matrix element of the
operator $dH(\lambda)/d\lambda$:
\eq
\frac{dE_n(\lambda)}{d\lambda}
=\biggl\langle\Psi_n(\lambda)\biggl|\frac{dH(\lambda)}{d\lambda}\biggr|\Psi_n(\lambda)\biggr\rangle\, .
\en
In the context of QCD, one often identifies the abstract parameter $\lambda$ with the
quark masses $m_q$ and studies the dependence of the hadron spectrum on the quark
masses. Since the quark mass dependent part of the QCD Hamiltonian takes the form
$H_m=\sum_qm_q\bar qq$, the dependence of, say, the nucleon mass on the quark masses 
is given by
\eq
\frac{dm_N}{dm_q}=\frac{1}{2m_N}\,\langle N|\bar qq|N\rangle\, .
\en
Here, the factor $1/(2m_N)$ emerges from the relativistic normalization of the
one-particle states, $\langle N'|N\rangle=(2\pi)^32E_N\delta^3({\bf p}'_N-{\bf p}^{}_N)$.

Below, we shall explicitly consider  only the three light quark flavors $q=u,d,s$ and, for
simplicity, assume that isospin is conserved: $m_u=m_d=\hat m$. The 
non-strange and strange $\sigma$-terms of the nucleon are defined, respectively, as:
\eq
\sigma_N=\frac{\hat m}{2m_N}\langle N|\bar uu+\bar dd|N\rangle\, ,\quad\quad
\sigma^s_N=\frac{m_s}{2m_N}\langle N|\bar ss|N\rangle\, ,
\en
and the strangeness content of the nucleon is given by
\eq\label{eq:strangeness}
y=\frac{2\langle N|\bar ss|N\rangle}{\langle N|\bar uu+\bar dd|N\rangle}\, .
\en
These $\sigma$-terms contain important information about the effect of the
explicit chiral symmetry breaking ($m_q\neq 0$) on the hadronic observables. 
In addition,  the nucleon $\sigma$-term is an important input for the 
estimates of  WIMP cross sections in dark matter direct detection experiments
see, e.g.,~\cite{Bottino:1999ei,Ellis:2008hf,Belanger:2008sj,Crivellin:2015bva,Hoferichter:2015ipa}, as well as in searches for the lepton flavor violation 
\cite{Cirigliano:2009bz,Crivellin:2014cta}
and electric dipole 
moments~\cite{Crewther:1979pi,Bsaisou:2012rg,Engel:2013lsa,deVries:2015gea,deVries:2015una}. 
The  strange $\sigma$-term of the nucleon
is relevant for the kaon condensation and the formation of the neutron stars, as well
as the study of the heavy ion collisions~\cite{Kaplan:1986yq,Nelson:1987dg,Meissner:2001gz}, etc. The extraction of the $\sigma$-terms from the experimental
data is a very delicate issue since, in particular, it 
implies an analytic continuation of the amplitudes below threshold 
(to the Cheng-Dashed point). In this respect, we note that
from a thorough theoretical analysis of the problem on the basis of dispersion relations
and using the input from Chiral Perturbation Theory (ChPT), 
in Ref.~\cite{Gasser:1990ce} the value $\sigma_N\simeq 45~\mbox{MeV}$ 
was obtained for the non-strange $\sigma$-term. The most recent and comprehensive
analysis, carried out in 
Ref.~\cite{Hoferichter:2015dsa}, is based on the Roy-Steiner equations for  $\pi N$ 
scattering and yields a larger value $\sigma_N=(59.1\pm 3.5)~\mbox{MeV}$.
As explained in detail in Ref.~\cite{Hoferichter:2015hva}, the bulk of the difference
can be traced back to the new and improved values of the pion-nucleon scattering lengths
deduced from pionic hydrogen and deuterium. This conclusion has been strengthened by a recent
reanalysis of the low-energy pion-nucleon scattering data \cite{RuizdeElvira:2017stg}.

In recent years, the $\sigma$-terms have been also measured on the 
lattice~\cite{Horsley:2011wr,Junnarkar:2013ac,Durr:2015dna,Yang:2015uis,Abdel-Rehim:2016won,Bali:2012qs,Bali:2016lvx}.
In general, two methods are employed in these measurements: a direct measurement
of the matrix element and extracting the $\sigma$-terms from the quark mass 
dependence of the hadron masses with the use of the Feynman-Hellmann theorem.
The results obtained with the use of both methods are compatible with each other.\footnote{Note that the relation of these extractions for the pion-nucleon $\sigma$-term to the Roy-Steiner analysis is discussed in Ref.~\cite{Hoferichter:2016ocj}.}
The above example is a nice demonstration of the fact that the lattice provides
us with additional tools to study the structure of hadrons: while the quark masses
on the lattice are free parameters, the experimental data correspond to the physical
values of the quark masses that can not be varied.

The strangeness content of the nucleon, $y_N$, which was defined in 
Eq.~(\ref{eq:strangeness}), is closely related to the $\sigma$-term and 
measures the contribution of the strange quarks to the nucleon mass.
It can be shown that a large magnitude of the non-strange $\sigma$-term might 
signal a large violation
of the Gell-Mann-Okubo rule for the baryon octet, and/or a large admixture
of the $\bar ss$ state to the physical nucleon state that, in its turn, would mean that the
properties of the nucleon must differ significantly from those obtained in a simple 
quark model.

In this paper, we shall demonstrate that the knowledge of the $\sigma$-terms allows
one to answer the question about the nature of the given hadronic states -- namely,
whether these states are standard quark-model states ($qqq$ for baryons, $\bar qq$
for mesons), or contain a sizable exotic admixture (pentaquarks, tetraquarks, $\ldots$).
This criterion, which is a straightforward extension of the argument with the strangeness
content of the nucleon, was first proposed 
in Ref.~\cite{Bernard:2010fp}. In the present article, 
we extend the discussion to the resonance case. 
It will be  shown that the 
quark mass dependence of the measured hadron masses in the multiplets 
obeys certain constraints, if these states are well described by the quark model. 
Very different constraints emerge, say, for  tetraquarks, pentaquarks, etc.
We verify this statement by explicit calculations in ChPT, as well as by using large-$N_c$
arguments. Note that, at present, a different strategy for identifying the exotic 
multiquark states is used on the lattice. Namely, one picks up a large set, including both
the standard three-quark (quark-antiquark) operators, as well as the multiquark operators,
and calculates the overlap of a given state with the states produced by these operators
from the vacuum. A large overlap with the multiquark state would then signal that
a given hadron contains a large exotic component and {\em vice versa.} Albeit intuitively
crystal clear, the argument has to be taken with a grain of salt. The overlap integrals contain
 information about the short-range physics (e.g., about the smearing used in the
construction of the operators) and are therefore not observable quantities. Consequently,
the statements about the nature of the states, which were made on the basis of the calculation 
of the overlaps, are not completely unambiguous. On the contrary, the measured masses
of the hadrons on the lattice are observable quantities and, thus, the constraints
on the quark mass dependence, considered
in the present article, are devoid of such short-distance ambiguities.
   
Furthermore, note that  all candidates for exotica in QCD are resonances. 
Consequently, we shall need to extend
the derivation of the Feynman-Hellmann theorem
to the unstable states in quantum field theory (note that, in Ref.~\cite{Ziesche:1986bp},
a generalization for the so-called Gamov states has been considered within the
framework of  non-relativistic quantum mechanics, see also Ref.~\cite{Tachibana}). 
This extension is the one of the main results
of the present article. Note also that it is not clear, how  the overlaps could be used
in this case -- even in principle. As it is well-known,  resonances do not 
correspond to a single
energy level on the lattice but rather can be associated to a group of a close-by levels.
Introducing a novel method to distinguish the exotic states
on the lattice seems to be inevitable from this viewpoint
as well.

The layout of the present paper is as follows. In section~\ref{sec:criterion}, we consider
the quark mass dependence in the flavor multiplets and derive the above-mentioned
constraints.  
Section~\ref{sec:verification} contains the discussion of this dependence within
ChPT and large-$N_c$ QCD. In section~\ref{sec:lattice}, we perform the
analysis of the available lattice data from the QCDSF collaboration
on the quark mass dependence of the low-lying mesons, baryon octet and baryon decuplet (only stable states)
and demonstrate that, as expected, these contain a reasonably
small admixture of exotica. 
Finally, in section~\ref{sec:FH}, the Feynman-Hellmann theorem 
for the resonance states is derived. 
Section~\ref{sec:concl} contains our conclusions. 

\section{Distinguishing the quark-model states from exotic states}
\label{sec:criterion}
 
The very notion of an exotic state in QCD needs the quark model as a reference point.
Below, we shall generalize the notion of the strangeness content of the nucleon and 
define  {\em observable} quantities, which characterize a given hadronic multiplet.
In the quark model, these quantities are given by exact group-theoretical factors. 
Then, the closeness of the measured values of these quantities to the 
quark-model values will be interpreted that the hadrons in this multiplet are
predominately non-exotic.

We explain the method for the example of the baryon octet.
Let $|B\rangle,~B=N,\Sigma,\Xi,\Lambda$, denote the eight different states of the octet. 
The Gell-Mann-Okubo type relations for the $\bar qq$ matrix elements  take the form
\eq
y_B&\doteq&\langle B|\bar ss|B\rangle=a_B+b_BY+c_B\biggl(I(I+1)-\frac{1}{4}\,Y^2-1\biggr)\, ,
\nonumber\\[2mm]
x_B&\doteq&\langle B|\bar uu+\bar dd|B\rangle=a_B'+b_B'Y+c_B'\biggl(I(I+1)-\frac{1}{4}\,Y^2-1\biggr)\, ,
\en
where $Y$ and $I$ stand, respectively, 
for the hypercharge and the isospin of the state $|B\rangle$. 
From the Feynman-Hellmann theorem, one gets
\eq
y_B=\frac{dm_B^2}{dm_s}\, ,\quad\quad x_B=\frac{dm_B^2}{d\hat m}\, .
\en
The following relations are straightforwardly obtained:
\eq
\gamma_B&\doteq&\frac{c_B}{a_B-c_B}= \frac{2y_\Sigma-y_N-y_\Xi}{2(y_N+y_\Xi)-y_\Sigma}\, ,
\nonumber\\[2mm]
\beta_B&\doteq&\frac{2b_B}{3(a_B-c_B)}= \frac{y_N-y_\Xi}{2(y_N+y_\Xi)-y_\Sigma}\, ,
\nonumber\\[2mm]
\gamma'_B&\doteq&\frac{c'_B}{a'_B-c'_B}= \frac{2x_\Sigma-x_N-x_\Xi}{2(x_N+x_\Xi)-x_\Sigma}\, ,
\nonumber\\[2mm]
\beta'_B&\doteq&\frac{2b'_B}{3(a'_B-c'_B)}= \frac{x_N-x_\Xi}{2(x_N+x_\Xi)-x_\Sigma}\, .
\en
Note that the quantities $\gamma_B,\beta_B,\gamma_B',\beta_B'$ are scale-invariant
in QCD and are therefore devoid of  short-range ambiguities.

Further, these ratios can be straightforwardly calculated in the quark model, where the
matrix element of the operator $\bar qq$ is merely given by the {\em total number} 
of quarks and antiquarks
with flavor $f$ present in the state $B$. A simple calculation gives:
\eq\label{eq:quarkB}
\gamma_B=\gamma'_B=0\, ,\quad\quad \beta_B=-\frac{2}{3}\, ,\quad\quad
\beta'_B=\frac{1}{3}\, .
\en
The same method can be applied, e.g., to the pseudoscalar meson octet. 
In this case, $\beta_P=\beta'_P=0$ from the beginning, and the remaining
coefficients are equal to
\eq\label{eq:gammaP}
\gamma_P&\doteq& \frac{2(y_\pi-y_K)}{4y_K-y_\pi}\, ,
\nonumber\\[2mm]
\gamma'_P&\doteq&\frac{2(x_\pi-x_K)}{4x_K-x_\pi}\, ,
\en
where
\eq\label{eq:xyP}
y_P=\frac{dM_P^2}{dm_s}\, ,\quad\quad x_P=\frac{dM_P^2}{d\hat m}\, ,\quad\quad
P=\pi,K\, .
\en
Note that the above relations remain 
valid even in the presence of the $\eta-\eta'$ mixing.
The matrix elements with $\eta,\eta'$ are merely
not present there. The quark-model values
of the above coefficients are easily calculated:
\eq
\gamma_P=-\frac{1}{2}\, ,\quad\quad \gamma'_P=1\, ,
\en
whereas, for example, using the wave functions for the tetraquark octet
from Ref.~\cite{Hooft:2008we}, one gets completely different values:
$\gamma_P=1,~\gamma'_P=-\frac{1}{5}$.

The generalization to the other multiplets is straightforward. Below, we present the 
formulae for the baryon decuplet (note that, using SU(3) symmetry, one may
rewrite the following formulae in different equivalent forms)
\eq
\gamma_\Delta=\frac{y_\Omega-y_{\Sigma^*}}{2y_{\Sigma^*}}\, ,\quad\quad
\gamma'_\Delta=\frac{x_\Omega-x_{\Sigma^*}}{2x_{\Sigma^*}}\, .
\en
The quark-model values are $\gamma_\Delta=1$ and $\gamma'_\Delta=-\frac{1}{2}$.

The argument then goes as follows. On the lattice, one may obtain the quark mass
dependence of the various members of the multiplets and thus extract the 
scale-independent quantities
$\gamma,\gamma',\beta,\beta'$ for a given multiplet. Any statistically 
significant deviation of these quantities from their quark model values can be interpreted
as the effect coming from the SU(3) breaking and/or from the significant admixture
of the non-quark-model states. Moreover, on the lattice one could determine the
derivatives of the hadron masses with respect to the quark masses in the vicinity
of the SU(3)-symmetric point $m_s=\hat m$ as well. On the basis of this analysis
one could then unambiguously judge about the exotic content of the multiplets
in the vicinity of this point.

Up to this point, there is little new information. The arguments like given above have 
been used in the past already. However, as mentioned in the introduction, one would
like to generalize these arguments to the case of unstable states, which do not 
correspond to a single energy level on the lattice. Note that the different levels may
have different quark mass dependence, so the quantities 
$\gamma,\gamma',\beta,\beta'$ cannot be defined unambiguously in this case.
However, prior to considering the problem of the resonances, we would like to
{\em validate} our arguments in the large-$N_c$ limit, as well as through  direct
calculations in ChPT.

\section{Validation of the method}
\label{sec:verification}

\subsection{Large-$N_c$}

In the context of QCD, the quark-model relations are reproduced in the limit 
$N_c\to\infty$ that corresponds to the quenching of the virtual quark loops in the path
integral. One immediately arrives at the answer in the meson sector. First of all, note
that $y_\pi=0$  and, consequently, $\gamma_P=-\frac{1}{2}$ (without assuming  
the exact SU(3) symmetry). Further, one gets $x_\pi=2x_K$ and the quark-model value
$\gamma'_P=1$ is reproduced.

The situation with the baryons is more subtle. Of course, in the {\em quenched theory,}
%considering the diagrams shown in Fig.~\ref{fig:quark}, 
one immediately gets
$y_N=0$ (without assuming  the exact SU(3) symmetry) and $y_\Xi=2y_\Sigma$,
$x_N=3x_\Xi,~x_\Sigma=2x_\Xi$, so that the quark model values
$\gamma_B=0,~\beta_B=-\frac{2}{3},~\gamma'_B=0,~\beta'_B=\frac{1}{3}$ are again
reproduced. However, {\em in the large-$N_c$ limit,} the baryons consist of
$N_c$ quarks and not just of three quarks -- so, the above arguments do not apply
straightforwardly. In fact, it was shown that, in this limit, baryons represent static
objects (their mass grows like $N_c$), which at leading order in $N_c$
can be described by using, e.g., a constituent quark 
model~\cite{Witten,KarlPaton,Buchmann,Lebed,Manohar}. 
For any value of $N_c$, the baryons,
containing $N_c$ quarks, belong to a completely symmetric irreducible representation
of the SU(6) spin-flavor group. Counterparts of the ``usual'' baryons 
$N,\Sigma,\Xi,\ldots$ are those members of the larger multiplets, which
have the same spin-flavor quantum numbers. For example, the proton has 
spin $s=s_z=\frac{1}{2}$, isospin $I=I_z=\frac{1}{2}$, hypercharge $Y=1$,
charge $Q=1$ and baryon number $B=1$ for any given $N_c$.    

Let us now obtain the values of $\gamma_B,~\beta_B,~\gamma'_B,~\beta'_B$ in the
large-$N_c$ limit. As known, for an arbitrary $N_c$, the generators corresponding
to the baryon number and the hypercharge are given by
\eq
\hat B=\frac{1}{N_c}\,\mbox{diag}(1,1,1)\, ,\quad\quad
\hat Y=\frac{1}{N_c}\,\mbox{diag}(1,1,1-N_c)\, .
\en
It is straightforward to check that the mass term in the Lagrangian is given by
\eq
{\cal L}_m=\hat m\bar\psi((N_c-1)\hat B+\hat Y)\psi+m_s\bar\psi(\hat B-\hat Y)\psi\, .
\en
From the above equation one may read off the values for $x_B,y_B$ (up to an overall
normalization factor)
\eq
x_N&=&(N_c-1)+1\, ,\quad
x_\Sigma=(N_c-1)\, ,\quad
x_\Xi=(N_c-1)-1\, ,
\nonumber\\[2mm]
y_N&=&0\, ,\quad
y_\Sigma=1\, ,\quad
y_\Xi=2\, ,
\en
and, finally,
\eq
\gamma_B=\gamma_B'=0\,,\quad\quad
\beta_B=-\frac{2}{3}\,,\quad\quad
\beta'_B=\frac{2}{3(N_c-1)}\, .
\en
As we see, the quantity $\beta_B$ stays finite in the large-$N_c$ limit, whereas
the quantities $\gamma_B,\gamma_B',\beta_B'$ all vanish in this limit. On the other 
hand, at $N_c=3$ the
quark model results are reproduced. Note also that our results are in complete 
agreement with Ref.~\cite{Jenkins} at the leading order in $1/N_c$.

%\begin{figure}[t]
%\begin{center}
%\includegraphics*[width=10.cm,angle=0]{quark.eps}
%\end{center}
%\caption{Quark diagrams that define the quantities  $y_\Sigma$ and $y_\Xi$ ($N_c=3$).
%$\ell$ denotes the $u,d$ quarks and the cross stands for the insertion of the operator
%$\bar ss$.}\label{fig:quark}
%\end{figure}

\subsection{Chiral Perturbation Theory}

\subsubsection{Goldstone boson octet}

At one loop, the pion and kaon masses in  3-flavor ChPT are given by 
(see, e.g.~\cite{Gasser:1984gg})
\eq
M_\pi^2&=&\sMpi\biggl\{1+\mu_\pi-\frac{1}{3}\,\mu_\eta+2\hat m K_3+K_4\biggr\}
\nonumber\\[2mm]
M_K^2&=&\sMK\biggr\{1+\frac{2}{3}\,\mu_\eta+(\hat m+m_s)K_3+K_4\biggr\},
\en
where
\eq
\mu_P&=&\frac{\sMP}{32\pi^2 F_0^2}\,\ln\frac{\sMP}{\mu^2}\, ,\quad\quad P=\pi,K,\eta\, ,
\nonumber\\[2mm]
\sMpi&=&2\hat mB_0\, ,\quad
\sMK=(\hat m+m_s)B_0\, ,\quad
\sMeta=\frac{2}{3}\,(\hat m+2m_s)B_0\, ,
\nonumber\\[2mm]
K_3&=&\frac{8B_0}{F_0^2}\,(2L^r_8-L^r_5)\, ,\quad\quad
K_4=(2\hat m+m_s)\frac{16B_0}{F_0^2}\,(2L^r_6-L^r_4)\, ,
\en
with $B_0,F_0,L_i^r$ the parameters of the ChPT Lagrangian,  
$\sMP$ are the pseudoscalar meson masses squared at leading order,
and $\mu$ denotes
the scale of  dimensional regularization. Calculating the
parameters $\gamma_P,\gamma'_P$ from these expressions, we get
\eq
\gamma_P&=&-\,\frac{1}{2}\,\biggl\{
1+\frac{\sMpi}{96\pi^2 F_0^2}\biggl(
\ln\frac{\sMeta}{\mu^2}+1\biggr)
-\frac{12 \sMpi}{F_0^2}\,(2L^r_6-L^r_4)\biggr\}\, ,
\nonumber\\[2mm]
\gamma'_P&=&1+\frac{3\sMpi}{16\pi^2F_0^2}\,
\ln\frac{\sMpi}{\mu^2}
-\frac{6\sMeta+\sMpi}{48\pi^2F_0^2}\ln\frac{\sMeta}{\mu^2}
+\frac{7\sMpi-3\sMeta}{96\pi^2F_0^2}
\nonumber\\[2mm]
&+&\frac{36(\sMpi-\sMeta)}{F_0^2}\,(2L^r_8-L^r_5)
-\frac{24(3\sMeta-\sMpi)}{F_0^2}\,(2L^r_6-L^r_4)\, .
\en
If we recall now that, in the large-$N_c$ limit, $F_0=O(N_c^{1/2})$, 
$L_5^r,L_8^r=O(N_c)$, $L_4^r,L_6^r=O(1)$ and the meson masses are of order~1 as well,
it is seen that $\gamma_P,\gamma'_P$ tend to the quark-model values in the limit
$N_c\to\infty$ and $m_s=\hat m$. Further, one could estimate the deviation of these
parameters in the real world from their exact quark-model values. To this end, we
replace $F_0$ by the pion decay constant $F_\pi=92.2~\mbox{MeV}$ 
and $\stackrel{\sf o}{M}_\pi$, $\stackrel{\sf o}{M}_K$ by the physical meson masses
($\stackrel{\sf o}{M}_\eta$ is determined from the Gell-Mann-Okubo relation).
Using the central values for the low-energy constants (LECs) $L_i^r$ from 
Ref.~\cite{Gasser:1984gg} at $\mu=770~\mbox{MeV}$ 
\eq
L_4^r=L_6^r=0\, ,\quad\quad L_5^r=2.2\cdot10^{-3}\, ,\quad\quad
L_8^r=1.1\cdot10^{-3}\, ,
\en
we get 
\eq
\gamma_P=-\frac{1}{2}\,(1+9\cdot10^{-4})\, ,\quad\quad
\gamma'_P=1+4\cdot 10^{-2}\, .
\en
In order to estimate the uncertainty, we present the results obtained by using the LECs
from Ref.~\cite{Bijnens:2014lea}. There are different sets of LECs in this paper. We use
the $O(p^4)$ fit from table 1:
$L_4^r=L_6^r=0,~L_5^r=1.2\cdot 10^{-3},~L_8^r=0.5\cdot 10^{-3}$, 
as well as the fit to lattice data:
$L_4^r=0.04\cdot 10^{-3},~L_6^r=0.07\cdot 10^{-3},~L_5^r=0.84\cdot 10^{-3},~L_8^r=0.36\cdot 10^{-3}$, 
see table 5 of this paper. The results are:
\eq
\gamma_P&=&-\frac{1}{2}\,(1+9\cdot10^{-4})\, ,\quad\quad
\gamma'_P=1+0.30 \quad\quad [O(p^4)]\, ,
\nonumber\\[2mm]
\gamma_P&=&-\frac{1}{2}\,(1-2\cdot10^{-3})\, ,\quad\quad
\gamma'_P=1-7\cdot 10^{-2} \quad\quad [\mbox{lattice}]\, ,
\en
The SU(3)-symmetric point is achieved from the physical point by varying the quark 
masses such that the sum of all quark masses $2\hat m+m_s=\mbox{const}$.
This corresponds to $M_\pi=M_K=M_\eta=413~\mbox{MeV}$. At this point, using the
LECs from Ref.~\cite{Gasser:1984gg}, we get 
\eq
\gamma_P=-\frac{1}{2}\,(1-5\cdot10^{-3})\, ,\quad\quad
\gamma'_P=1-2\cdot 10^{-2}\, ,
\en
whereas the the use of the LECs from Ref.~\cite{Bijnens:2014lea} gives the following
answer
\eq
\gamma_P&=&-\frac{1}{2}\,(1-5\cdot10^{-3})\, ,\quad\quad
\gamma'_P=1-2\cdot 10^{-2} \quad\quad [O(p^4)]\, ,
\nonumber\\[2mm]
\gamma_P&=&-\frac{1}{2}\,(1-3\cdot10^{-2})\, ,\quad\quad
\gamma'_P=1-0.12 \quad\quad [\mbox{lattice}]\, ,
\en
As we see, the corrections to the quark model values for the pseudoscalar octet
are reasonably small. The uncertainties, which stem from the LECs, are however sizable
and the correlations between various LECs should be taken into account.

\subsubsection{Ground-state baryon octet}

The case of the baryon octet is another illustration of the method. One could use the result
on the baryon masses in ChPT, which are available in the literature,
see e.g. Refs.~\cite{Bernard:1993nj,Borasoy:1996bx,Frink:2004ic}. In this
article, we restrict ourselves to the $O(p^2)$ calculations. The relevant part of
the effective Lagrangian is given by
\eq\label{eq:L2}
{\cal L}^{(2)}=b_D\langle \bar B\{\chi_+,B\}\rangle
+b_F\langle \bar B[\chi_+,B]\rangle
+b_0\langle \bar BB\rangle\langle\chi_+\rangle+\cdots\, ,
\en
where the matrix $B$ is the baryon octet field and $\chi_+=u^\dagger\chi u^\dagger+u\chi u$, 
with $\chi=2B_0{\cal M}\,,~{\cal M}=\mbox{diag}(\hat m,\hat m,m_s)$. Further,
$u=\exp(i\phi/(2F_0))$ is the Goldstone boson field and $b_D,b_F,b_0$ are the
pertinent LECs. The expressions for the coefficients $\gamma_B,\gamma'_B,\beta_B,\beta'_B$ 
can be readily obtained from the calculated baryon masses at this order
\eq\label{eq:BChPT}
\gamma_B&=&-\frac{2b_D}{3b_0+4b_D}\, ,\quad\quad
\gamma'_B=\frac{b_D}{3b_0+b_D}\, ,
\nonumber\\[2mm]
\beta_B&=&-\frac{2b_F}{3b_0+4b_D}\, ,\quad\quad
\beta'_B=\frac{b_F}{3b_0+b_D}\, .
\en
Unlike the Goldstone boson case, these expressions do not reduce automatically
at lowest order to the quark-model values given by Eq.~(\ref{eq:quarkB}). 
The quark model results are reproduced, if $b_D/b_0=0$ and $b_F/b_0=1$.
This statement seems to be supported by
phenomenological values of these LECs~\cite{Bernard:1993nj} (all in $\mbox{GeV}^{-1}$ units)
\eq
-0.79\leq b_0\leq -0.70\, ,\quad\quad
0.01\leq b_D\leq 0.07\, ,\quad\quad
-0.61\leq b_F\leq -0.48\, .
\en
It could be interesting to consider
the constraints on the higher-order LECs, which emerge in a similar fashion 
and, possibly, from
the observables other than the baryon masses as well. However, since the convergence
in the SU(3) baryon ChPT is a somewhat  painful issue~\cite{Hoferichter:2015tha,Siemens:2016jwj}, we do not consider any 
further constraints here. 
The results of Ref.~\cite{Ren:2015dvx} support this statement. For example,
the NLO results for these constants are: $b_0=-0.273(6),~b_D=0.0506(17),~b_F=-0.179(1)$, whereas at the NNLO one has: $b_0=-0.886(5),~b_D=0.0482(17),~b_F=-0.517(7)$
(all in $\mbox{GeV}^{-1}$ units). One sees that the general pattern persists, albeit the
values of the LECs vary quite a bit.

In Ref.~\cite{Jenkins}, comparing the chiral Lagrangians at arbitrary $N_c$ and at
$N_c=3$, it was shown that $b_D/b_0=0$ and $b_F/b_0=1$ hold in the limit of large
$N_c$. Therefore, the quark-model values are exact at $N_c=3$ up to the 
corrections  that vanish in the large-$N_c$ limit. Note also that in Eq.~(\ref{eq:BChPT})
the large-$N_c$ limit can not be performed straightforwardly, because it is
obtained from the Lagrangian given in Eq.~(\ref{eq:L2}), which is
 defined at $N_c=3$ only.
 In fact, as easily seen, the above constraints on $b_0,b_D,b_F$ are incompatible
with $\beta'_B\to 0$ and $\beta_B\neq 0$ in the large-$N_c$ limit.

\subsubsection{Ground-state vector meson octet}

The vector meson octet contains unstable particles. So, strictly speaking, our 
formulas are not directly applicable there. However, in order to get an intuitive
understanding of the problem, one could still use an effective Lagrangian with
vector mesons (see, e.g., Ref.~\cite{Bijnens:1997ni}) and evaluate the pertinent
coefficients $\gamma_V,\gamma'_V$, defined through the Eqs.~(\ref{eq:gammaP},\ref{eq:xyP}) 
with the replacements $\pi\to\rho$ and $K\to K^*$. The quark
mass dependent part of the effective Lagrangian is given by~\cite{Bijnens:1997ni}:
\eq
{\cal L}_m&=&a_1\langle\{W_\mu^\dagger,W^\mu\}\chi_+\rangle
+a_{13}\langle W_\mu^\dagger W^\mu\rangle\langle\chi_+\rangle
\nonumber\\[2mm]
&+&c_1\langle W_\mu^\dagger\chi_+W^\mu\chi_+\rangle
+c_2\langle\{W_\mu^\dagger,W^\mu\}\chi_+\chi_+\rangle\, ,
\en
where the LEC $a_{13}$ is suppressed at $O(N_c^{-1})$ with respect to the other LECs
(we do not display the $1/N_c$-suppressed terms at order $p^4$).

As it is known, the Goldstone boson loops are suppressed by a factor $N_c$. For this reason,
we drop them in the expression of the masses completely. The contact term 
contribution to the masses is given below:
\eq
M_\rho^2&=&\sMV+8a_1B_0\hat m
+4a_{13}B_0(2\hat m+m_s)+16(c_1+2c_2)(B_0\hat m)^2\, ,
\nonumber\\[2mm]
M_{K^*}^2&=&\sMV+4a_1B_0(\hat m+m_s)+4a_{13}B_0(2\hat m+m_s)
\nonumber\\[2mm]
&+&16c_1B_0^2\hat mm_s+16c_2B_0^2(\hat m^2+\hat m_s^2)\, ,
\en
with $\sMV$ the vector meson octet mass in the chiral limit.
From this, one can immediately read off the expressions for $\gamma_V,\gamma'_V$:
\eq
\gamma_V&=&\frac{2(y_\rho-y_{K^*})}{4y_{K^*}-y_\rho}=
-\frac{1}{2}\, \frac{4a_1}{4a_1+3a_{13}}+O(p^4)+O(p^2N_c^{-1})\, ,
\nonumber\\[2mm]
\gamma'_V&=&\frac{2(x_\rho-x_{K^*})}{4x_{K^*}-x_\rho}=
\frac{a_1}{a_1+3a_{13}}-\frac{12c_1(M_K^2-M_\pi^2)}{a_1+a_{13}}+O(p^4)
+O(p^2N_c^{-1})\, .\hspace*{.6cm}
\en
It is seen that the corrections to $\gamma_V,\gamma'_V$ 
vanish in the large-$N_c$ and SU(3) symmetry limit. 
From this example, it becomes clear, how the effective
theory for the exotic multiplets (e.g., the scalar mesons) could be constructed. The 
operator 
basis, whose form is dictated by the symmetries, is the same in case of the standard
and exotic particles. The difference emerges at the level of the LECs: Certain LECs in the
effective Lagrangian with exotic particles are not suppressed in the large-$N_c$ limit.

\section{Testing with lattice data}
\label{sec:lattice}

The QCDSF collaboration has studied the quark mass 
dependence of hadron masses in various meson and baryon multiplets. 
This analysis provides us with an ideal input to test our predictions.
In order to make the comparison straightforward, both for the bare and renormalized 
quark masses, we define
\eq
\bar m&=&\frac{1}{3}\,(2\hat m+m_s)\, ,\quad\quad m_1=\hat m-m_s\, ,
\nonumber\\[2mm]
\bar m^r&=&\frac{1}{3}\,(2\hat m^r+m_s^r)\, ,\quad\quad m_1^r=\hat m^r-m_s^r\, ,
\en
The bare and renormalized quark masses are related by
\eq
\bar m^r=Z_m^S\bar m\, ,\quad
m_1^r=Z_m^{NS}m_1\, ,\quad\quad
Z_m^S/Z_m^{NS}=1+\alpha_Z\, .
\en
Calculating the derivatives with respect to the renormalized masses, we get
\eq
y_a&\doteq&\frac{\partial M_a^2}{\partial m_s^r}\biggr|_{\hat m^r={\sf const}}
=\frac{1}{3}\,\frac{\partial M_a^2}{\partial \bar m^r}\biggr|_{m_1^r={\sf const}}
-\frac{\partial M_a^2}{\partial m_1^r}\biggr|_{\bar m^r={\sf const}}\, ,
\nonumber\\[2mm]
x_a&\doteq&\frac{\partial M_a^2}{\partial \hat m^r}\biggr|_{m_s^r={\sf const}}
=\frac{2}{3}\,\frac{\partial M_a^2}{\partial \bar m^r}\biggr|_{m_1^r={\sf const}}
+\frac{\partial M_a^2}{\partial m_1^r}\biggr|_{\bar m^r={\sf const}}\, .
\en
Here, $M_a$ denote the hadron masses in a given multiplet.
One may relate this to the derivative with respect to the bare masses
\eq
\frac{\partial M_a^2}{\partial \bar m^r}
&=&\frac{1}{Z_m^S}\,\frac{\partial M_a^2}{\partial \bar m}
=\frac{1}{Z_m^{NS}(1+\alpha_Z)}\,\frac{\partial M_a^2}{\partial \bar m}\, ,
\nonumber\\[2mm]
\frac{\partial M_a^2}{\partial \bar m_1^r}
&=&\frac{1}{Z_m^{NS}}\,\frac{\partial M_a^2}{\partial m_1}\, .
\en
SU(3) symmetry in the vicinity of the symmetric point introduces
additional constraints on the derivatives with respect to $m_1$. Namely, retaining
only linear terms in $m_1$, the hadron masses in the multiplets are given by~\cite{Bietenholz:2011qq}

\bigskip

{\em pseudoscalar octet:}

\eq
M_\pi^2=M_P^2+\frac{2}{3}\,\alpha_P m_1\, ,\quad
M_K^2=M_P^2-\frac{1}{3}\,\alpha_P m_1\, ,\quad
M_{\eta_s}^2=M_P^2-\frac{4}{3}\,\alpha_P m_1\, .
\en

\bigskip

{\em vector octet:}

\eq\label{eq:V}
M_\rho=M_V+\frac{2}{3}\,\alpha_V m_1\, ,\quad
M_{K^*}=M_V-\frac{1}{3}\,\alpha_V m_1\, ,\quad
M_{\phi_s}=M_V-\frac{4}{3}\,\alpha_V m_1\, .
\en

\bigskip

{\em baryon octet:}

\eq\label{eq:B}
M_N&=&M_B+A_1m_1\, ,\quad
M_\Lambda=M_B+A_2m_1\, ,
\nonumber\\[2mm]
M_\Sigma&=&M_B-A_2m_1\, ,\quad
M_\Xi=M_B-(A_1-A_2)m_1\, .
\en

\bigskip

{\em baryon decuplet:}

\eq\label{eq:DD}
M_\Delta&=&M_D+Am_1\, ,\quad
M_{\Sigma^*}=M_D\, ,
\nonumber\\[2mm]
M_{\Xi^*}&=&M_D-Am_1\, ,\quad
M_\Omega=M_D-2m_1\, .
\en

The expansion coefficients depend on the variable $\bar m$. Calculating the derivatives
with respect to the variables $\bar m,\,m_1$ and introducing the notation
\eq
\lambda_P=\frac{1}{1+\alpha_Z}\,\frac{dM_P^2}{d\bar m}\, ,\quad\quad
\lambda_A=\frac{1}{1+\alpha_Z}\,\frac{dM_A}{d\bar m}\, ,\quad A=V,B,D\, ,
\en
one gets
\eq
\gamma_P&=&\frac{-2(\alpha_P/\lambda_P)}{1+2(\alpha_P/\lambda_P)}\, ,\quad\quad
\gamma'_P=\frac{(\alpha_P/\lambda_P)}{1-(\alpha_P/\lambda_P)}\, ,
\nonumber\\[2mm]
\gamma_V&=&\frac{-2(\alpha_V/\lambda_V)}{1+2(\alpha_V/\lambda_V)}\, ,\quad\quad
\gamma'_V=\frac{(\alpha_V/\lambda_V)}{1-(\alpha_V/\lambda_V)}\, ,
\nonumber\\[2mm]
\gamma_B&=&\frac{(3A_2/\lambda_B)}{1-(3A_2/\lambda_B)}\, ,\quad\quad
\gamma'_B=\frac{-(3A_2/\lambda_B)/2}{1-(3A_2/\lambda_B)/2}\, ,
\nonumber\\[2mm]
\beta_B&=&\frac{-2A_1/\lambda_B+A_2/\lambda_B}{1-(3A_2/\lambda_B)}\, ,\quad\quad
\beta'_B=\frac{A_1/\lambda_B-A_2/(2\lambda_B)}{1+(3A_2/\lambda_B)/2}\, ,
\nonumber\\[2mm]
\gamma_D&=&\frac{3A}{\lambda_D}\, ,\quad\quad\gamma'_D=-\frac{3A}{2\lambda_D}\, .
\en
Note that the above relations are valid exactly in the SU(3)-symmetric point, where $m_1=0$ and the masses of all hadrons in the same multiplet are equal.

The quark-model values are
\eq\label{ratios-su3}
\frac{2\alpha_P}{\lambda_P}=\frac{2\alpha_V}{\lambda_V}=1\, ,\quad\quad
\frac{3A_1}{\lambda_B}=1\, ,\quad\quad
\frac{3A_2}{\lambda_B}=0\, ,\quad\quad
\frac{3A}{\lambda_D}=1\, ,
\en
whereas, e.g., the tetraquark value is completely different:
\eq
\frac{2\alpha_T}{\lambda_T}=-\frac{1}{2}\, .
\en
That is, even the sign is different from the case of the the ordinary $\bar qq$-mesons, 
allowing one to clearly distinguish between both QCD configurations.

Having set the definitions, we proceed with the verification of our method using the lattice data from Ref.~\cite{Bietenholz:2011qq}.
We will only consider results corresponding to the largest lattice size $32^3\times 64$, 
hence avoiding the discussion of finite-size lattice corrections. 
Bare quark masses are defined as
\begin{equation}\label{eq:mq}
a m_q=\frac{1}{2}\left(\frac{1}{\kappa_{q}}-\frac{1}{\kappa_{0,c}}\right).
\end{equation}

On the symmetric line, i.e. for $\hat m=m_s=\bar m$ and $\kappa_l=\kappa_s=\kappa_0$, the QCDSF results for the
pseudoscalars, vectors, octet and decuplet baryons are given in Table~\ref{tab:symm-line}.
Note that the results corresponding to the second row, $\kappa_0=0.12092$, have been slightly updated with respect to the values in Ref.~\cite{Bietenholz:2011qq}.
\begin{table}[t]
  \renewcommand{\arraystretch}{1.3}
  \centering
  \begin{tabular}{cccccc}\toprule
    $\kappa_0$& $a M_\pi$& $a M_\rho$& $a M_N$& $a M_\Delta$\\\midrule
    $0.12090$&$0.1747(5)$&$0.3341(34)$&$0.4673(27)$&$0.5675(64)$\\
    $0.12092^*$&$0.1647(4)$&$0.3282(41)$&$0.4443(59)$&$0.5577(112)$\\
    $0.12095$&$0.1508(4)$&$0.3209(27)$&$0.4329(49)$&$0.5541(80)$\\
    $0.12099$& $0.1285(7)$&$0.3006(59)$ &$0.4107(89)$&$0.5183(157)$
    \\\bottomrule
\end{tabular}
\caption{Updated values of hadron masses on the symmetric line from the $32^3\times 64$ lattice~\cite{Bietenholz:2011qq}. The input in the second raw is updated as
compared to Ref.~\cite{Bietenholz:2011qq} (this is marked by the $^*$).}
\label{tab:symm-line}
\end{table}

Chiral symmetry requires the pion mass to vanish in the chiral $\bar m\to 0$ limit, 
which allows one to determine the critical hopping parameter $\kappa_{0,c}$ by extrapolating the pseudoscalar masses through the symmetric line.  
Thus, considering at first order
\begin{equation}
M_\pi^2= \frac{a_\pi}{\kappa_0}+b_\pi=a_\pi\left(\frac{1}{\kappa_0}-\frac{1}{\kappa_{0,c}}\right)=2 a_\pi \bar m,\quad\kappa_{0,c}=-\frac{a_\pi}{b_\pi}, 
\end{equation}
and fitting the data in Table~\ref{tab:symm-line}, one can directly extract the value of the $a_\pi$ and $b_\pi$ coefficients, namely
\begin{align}\label{eq:symm-results}
&a_\pi=\frac{1}{2}\frac{dM_P^2}{d\bar m}=2.249\pm 0.037,& b_\pi=-18.57\pm0.31,
\end{align} 
so that the critical hopping parameter reads
\begin{equation}\label{eq:k0c-value}
\kappa_{0,c}=0.12098\pm 0.00283.
\end{equation}
The uncertainties have been computed using a bootstrap with a normally distributed sample of 1000 points, with a standard deviation defined from the 68\% of the distribution.

Once the critical hopping parameter $\kappa_{0,c}$ has been obtained, Eq.~\eqref{eq:mq} allows one to study the behavior of the vector and octet and decuplet baryons at the symmetric line.  
Considering again a linear fit to the data in Table~\ref{tab:symm-line}, we obtain
\begin{equation}\label{symm-results}
\frac{dM_V}{d\bar m}=7.66\pm 2.47,\quad \frac{dM_B}{d\bar m}=19.33\pm 2.3,\quad \frac{dM_D}{d\bar m}=12.28\pm 4.13,
\end{equation}
where the errors have been computed using again a bootstrap method.
\begin{figure}[t]
\begin{center}
\includegraphics*[width=10.cm,angle=0]{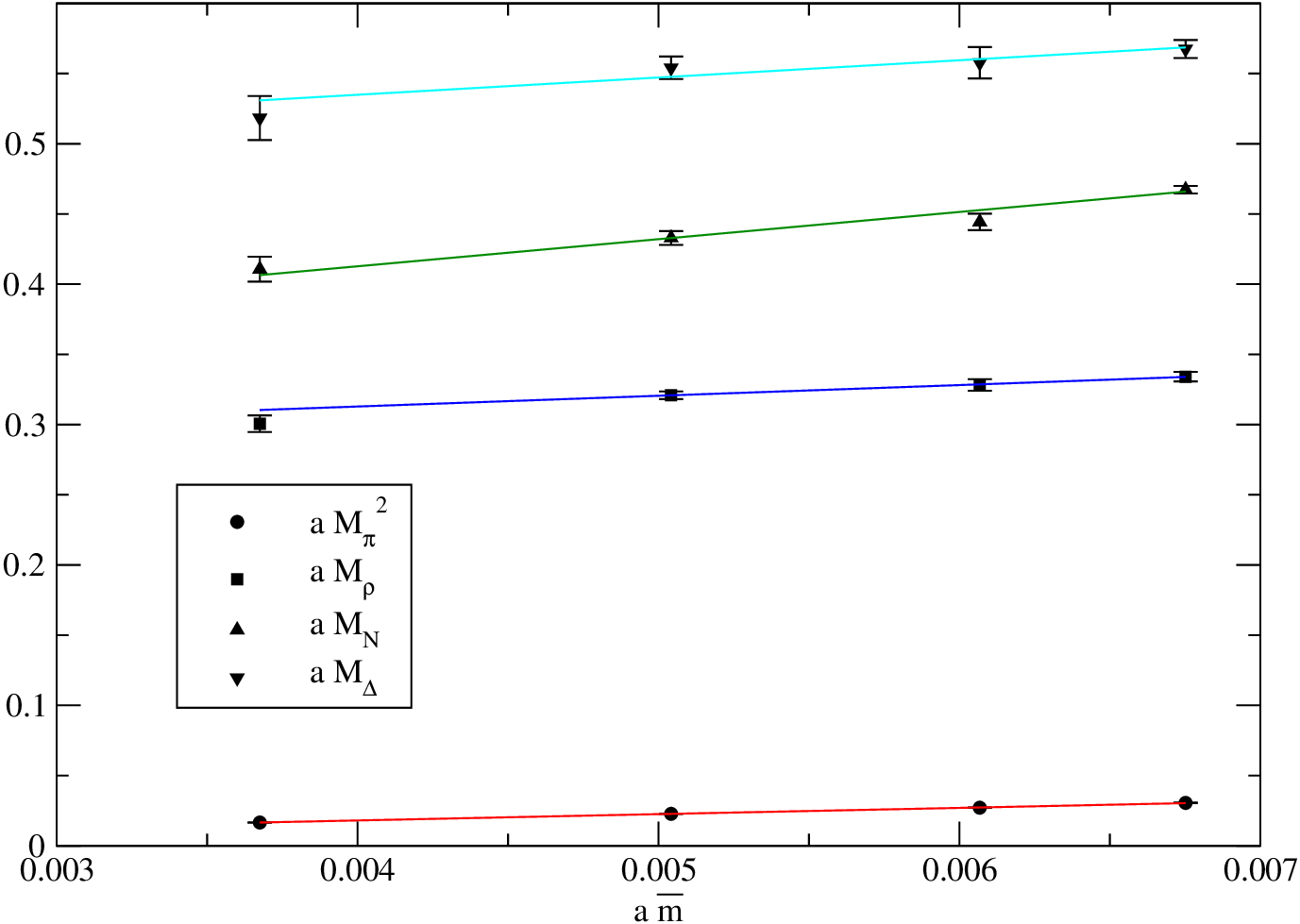}
\end{center}
\caption{Linear fit for pseudoscalar and vector mesons and octet and decuplet baryon along the symmetric line. In the case of the vectors, only the first three points have been considered in the fit, since the last one corresponds to an unstable particle.}\label{fig:symm-fit}
\end{figure}
The fit results, together with the lattice data in Table~\ref{tab:symm-line} are plotted in Fig.~\ref{fig:symm-fit}, showing a nice linear behavior and hence the fit quality. 
Note that for the vector mesons we have got rid of the last data point, which lies quite well above the $2m_\pi$ threshold and hence corresponds to an unstable particle.

Beyond the symmetric line, no reference to $\kappa_{0,c}$ is needed, since
\begin{equation}\label{eq:m1}
a m_1=\frac{1}{2}\left(\frac{1}{\kappa_l}-\frac{1}{\kappa_s}\right).
\end{equation}

The QCDSF collaboration results for pseudoscalar and vector mesons, and octet and decuplet baryons corresponding to the $32^3\times 64$ lattice size can be found in Tables~\ref{Tab:no-symm-M}-\ref{Tab:no-symm-B}.
Once more, the results of the fourth raw, i.e. for $\kappa_l=0.121145$ and $\kappa_s=0.120413$, are also updated as compared with Ref.~\cite{Bietenholz:2011qq}.
\begin{table}[t]
  \renewcommand{\arraystretch}{1.3}
  \centering
  \begin{tabular}{cccc}\toprule
    $(\kappa_l,\kappa_s)$&$aM_\pi$&$aM_K$&$aM_{\eta_s}$\\\midrule
    $(0.120900, 0.120900)$ & $0.1747(5)$  & $0.1747(5)$  & $0.1747(5)$  \\
    $(0.121040, 0.120620)$ & $0.1349(5)$  & $0.1897(4)$  & $0.2321(3)$  \\
    $(0.121095, 0.120512)$ & $0.1162(8)$  & $0.1956(5)$  & $0.2512(3)$  \\
    $(0.121145, 0.120413)^*$ & $0.09687(84)$&$0.2015(4)$  & $0.2682(3)$  \\\bottomrule
  \end{tabular}\\
\vspace{0.9cm}
  \begin{tabular}{cccc}\toprule
    $(\kappa_l,\kappa_s)$&$aM_\rho$&$aM_{K^*}$&$aM_{\phi_s}$\\\midrule
    $(0.120900, 0.120900)$ & $0.3341(34)$ & $0.3341(34)$ & $0.3341(34)$ \\
    $(0.121040, 0.120620)$ & $0.3127(38)$ & $0.3380(21)$ & $0.3632(14)$ \\
    $(0.121095, 0.120512)$ & $0.3123(43)$ & $0.3426(20)$ & $0.3738(11)$ \\
    $(0.121145, 0.120413)^*$ & $0.3227(65)$ & $0.3490(22)$ & $0.3874(11)$ \\\bottomrule    
  \end{tabular}
\caption{Pseudoscalar meson masses $aM_\pi$, $aM_K$ and $aM_{\eta_s}$ and vector meson masses $M_\rho$, $aM_{K^*}$ and $aM_{\phi_s}$ on the $\bar{m}=\mbox{constant}$ line from the $32^3\times 64$ lattice~\cite{Bietenholz:2011qq}.
The input in the fourth raw is updated as compared to Ref.~\cite{Bietenholz:2011qq} (this is marked by the $^*$)}
\label{Tab:no-symm-M}
\end{table}

\begin{table}[ht]
  \renewcommand{\arraystretch}{1.3}
  \centering
  \begin{tabular}{ccccc}\toprule
    $(\kappa_l,\kappa_s)$&$aM_N$&$aM_\Lambda$&$aM_\Sigma$&$aM_\Xi$\\\midrule
    $(0.120900, 0.120900)$ & $0.4673(27)$ & $0.4673(27)$ & $0.4673(27)$ & $0.4673(27)$ \\
    $(0.121040, 0.120620)$ & $0.4267(50)$ & $0.4547(43)$ & $0.4697(33)$ & $0.4907(21)$ \\
    $(0.121095, 0.120512)$ & $0.4140(61)$ & $0.4510(58)$ & $0.4690(37)$ & $0.4971(21)$ \\
    $(0.121145, 0.120413)^*$ & $0.3950(114)$ & $0.4460(65)$ & $0.4739(42)$ & $0.5073(21)$ \\\bottomrule    
  \end{tabular}\\
\vspace{0.9cm}
  \begin{tabular}{ccccc}\toprule
    $(\kappa_l,\kappa_s)$&$aM_\Delta$&$aM_{\Sigma^*}$&$aM_{\Xi^*}$&$aM_\Omega$\\\midrule
  $(0.120900, 0.120900)$ & $0.5675(64))$& $0.5675(64) $& $0.5675(64)$ & $0.5675(64)$ \\
  $(0.121040, 0.120620)$ & $0.5520(79) $& $0.5744(48) $& $0.5968(34)$ & $0.6194(28)$ \\
  $(0.121095, 0.120512)$ & $0.5161(185)$& $0.5541(98) $& $0.5812(52)$ & $0.6104(33)$ \\
  $(0.121145, 0.120413)^*$ & $0.4808(320)$& $0.5441(136)$& $0.5956(68)$ & $0.6382(35)$ \\
  \bottomrule    
  \end{tabular}
\caption{Octet baryon masses $aM_N$, $aM_\Lambda$, $aM_\Sigma$ and $aM_\Xi$ and decuplet baryon masses $M_\Delta$, $aM_{\Sigma^*}$, $aM_{\Xi^*}$ and $aM_\Omega$ on the $\bar{m}=\mbox{constant}$ line for the $32^3\times 64$ lattice~\cite{Bietenholz:2011qq}. The input in the fourth raw is updated as compared to Ref.~\cite{Bietenholz:2011qq} (this is marked by the $^*$)}
\label{Tab:no-symm-B}
\end{table}

\begin{figure}[t]
\begin{center}
\includegraphics*[width=7.2cm,angle=0]{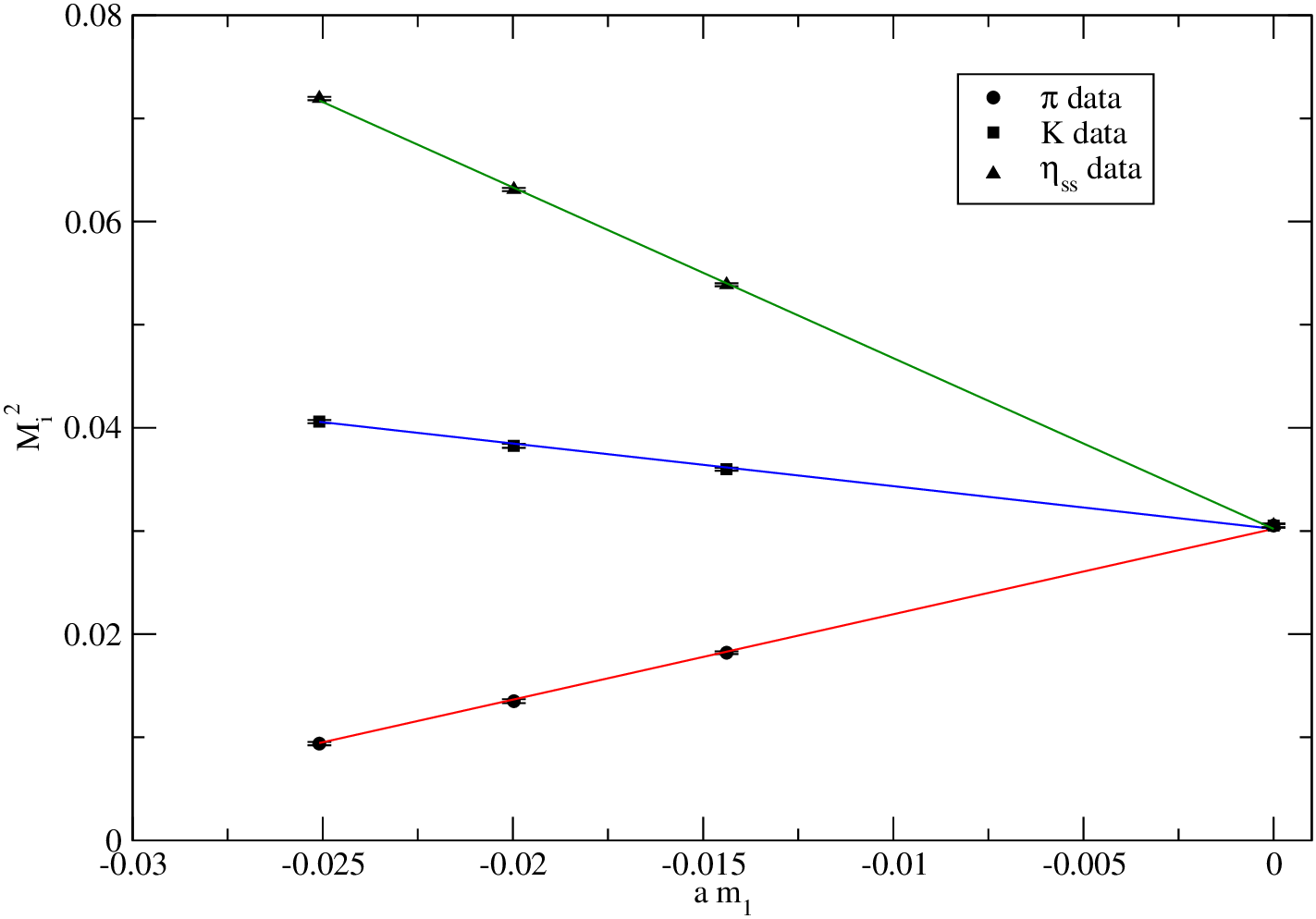}\includegraphics*[width=7.2cm,angle=0]{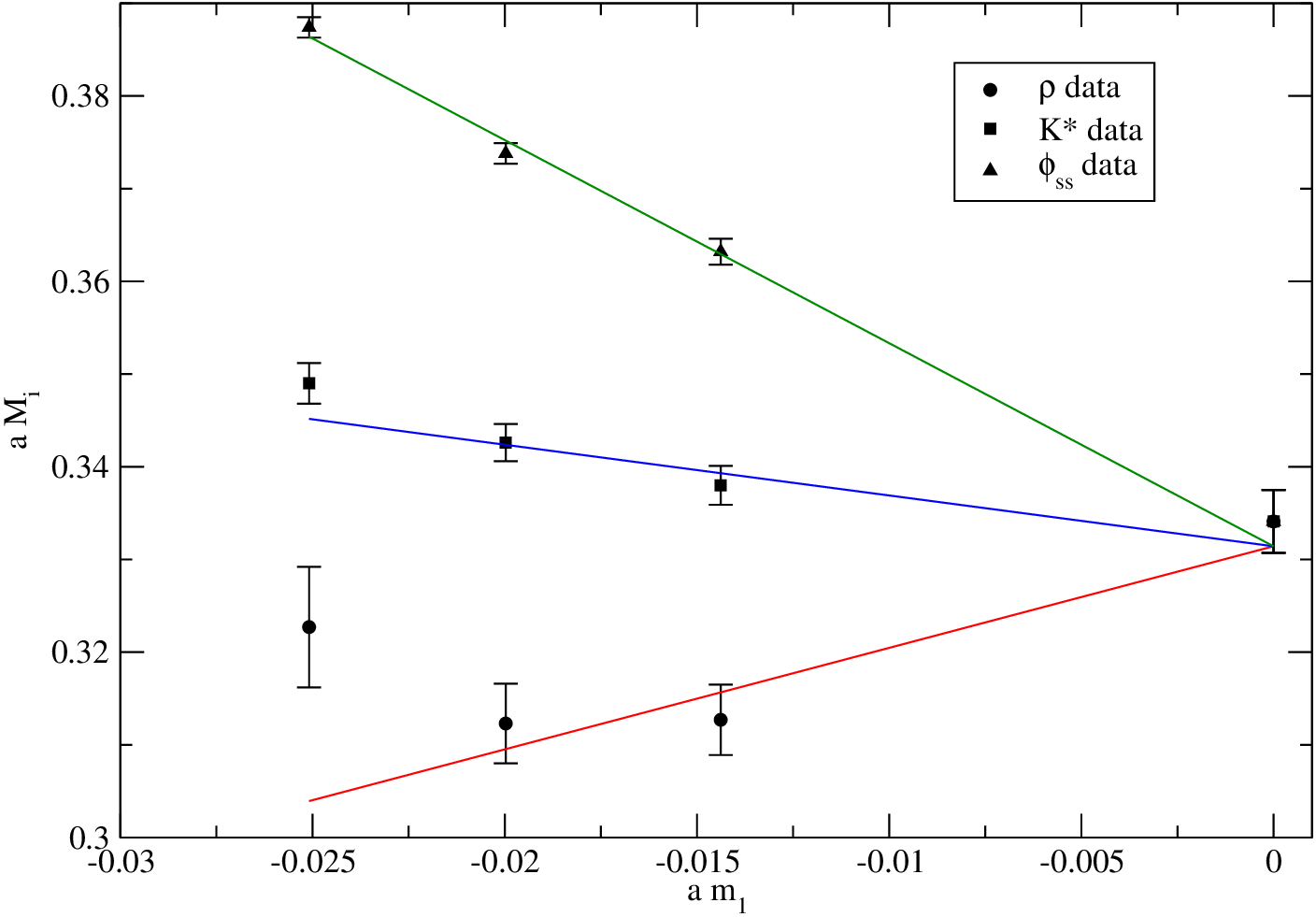}\\
\includegraphics*[width=7.2cm,angle=0]{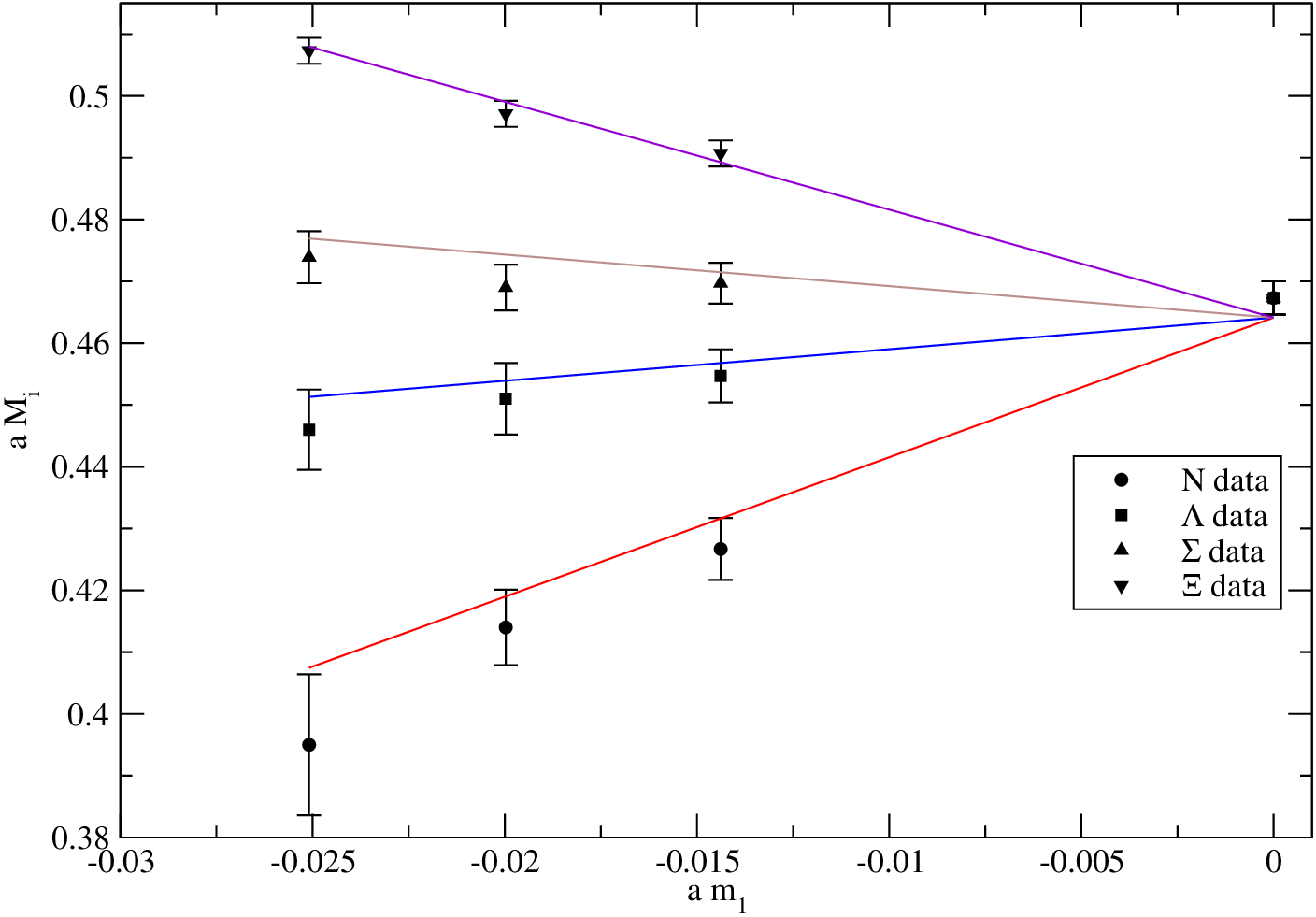}\includegraphics*[width=7.2cm,angle=0]{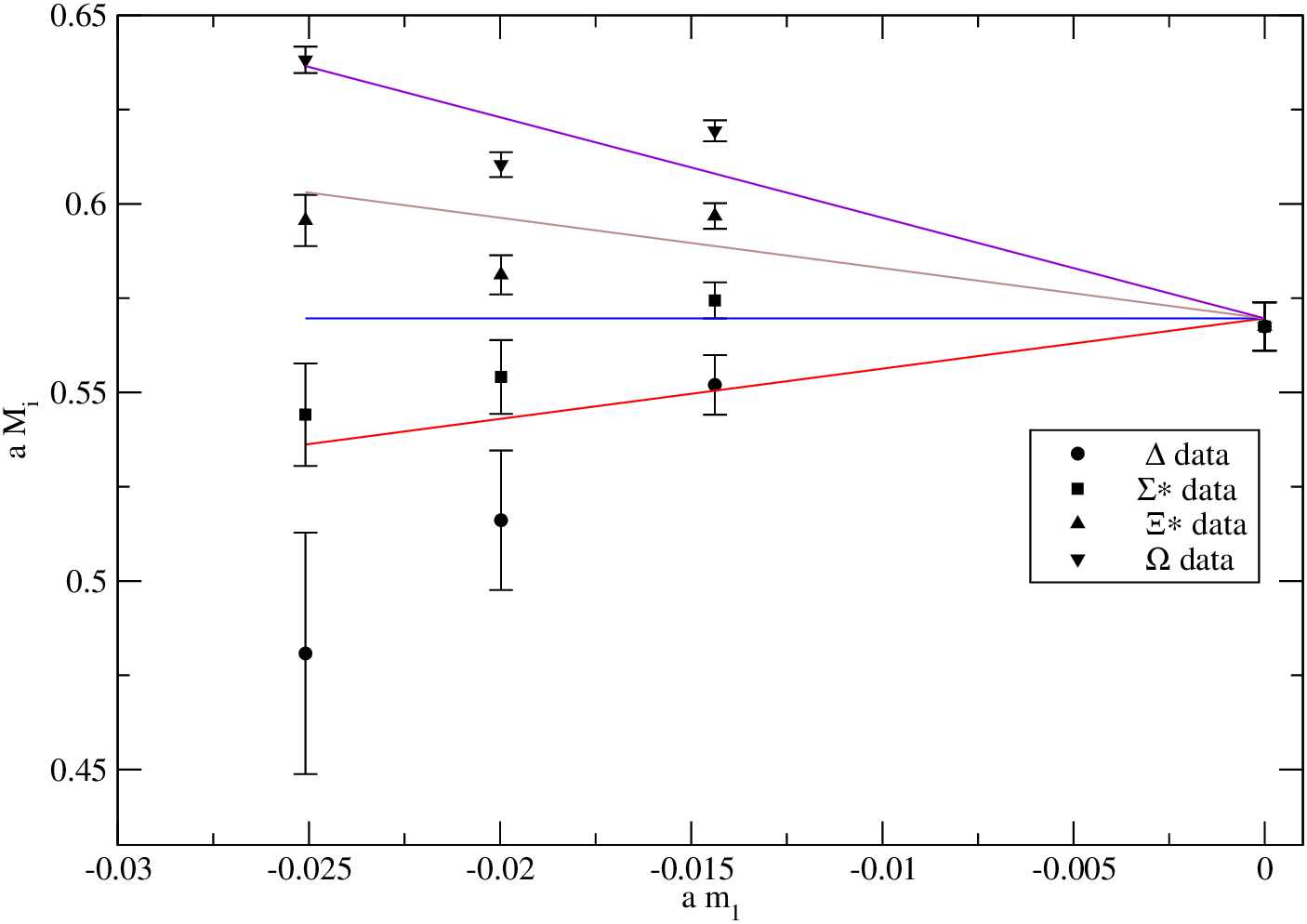}\\
\end{center}
\caption{Fit results for pseudoscalar (left up panel)
and vector (right up panel) mesons and octet (left down panel)
and decuplet (right down panel) baryons along the non-symmetric line.}\label{fig:non-symm}
\end{figure}
Using these data together to Eq.~\eqref{eq:m1}, we can perform a linear fit to compute the coefficients in Eqs.~\eqref{eq:V}-\eqref{eq:DD}, getting
\begin{align}\label{non-symm-results}
&\alpha_P=1.242\pm0.003,&\alpha_V=1.64 \pm0.05,&\nonumber\\
&A_1=2.26\pm0.09,& A_2=0.51\pm0.05,\\ 
&A=1.33\pm0.07,\nonumber
\end{align}
where, once more, the errors have been computed using a bootstrap with a normally distributed sample of 1000 points,
and the fit results have been plotted in Fig.~\ref{fig:non-symm}.
As in the symmetric case, we have not included the last two points for the $\rho$ and $K^*$ mesons, 
which correspond to mass values well above the $\pi\pi$ and $\pi K$ threshold, respectively, and hence to unstable particles.  
As one see in Fig.~\ref{fig:non-symm}, the linear fits work extremely well for pseudoscalar mesons, and relatively well for vectors and octet baryons. 
On the contrary, the decuplet results show deviations from a linear behavior. 
Nevertheless, their description does not improve with a quadratic function either and a linear behavior will be assumed in order to test our criterion for the decuplet.

\subsubsection*{Pseudoscalar mesons:}

Since the data on the pseudoscalar meson masses are almost linear in the quark masses -- in other words, the Gell-Mann-Oakes-Renner relation is obeyed very well,
we will assume an exact linear dependence to extract the renormalization constant $\alpha_Z$,
\begin{equation}\label{az-value}
\alpha_Z=\frac{1}{2\alpha_P}\frac{dM_P^2}{d\bar m}-1=0.811\pm0.029,
\end{equation}
so our criterion turns into an identity for the pseudoscalar mesons
\eq
\frac{2\alpha_P}{\lambda_P}=1.
\en

In the case of the vector mesons as well as for the octet and decuplet baryons, using the results in Eqs.~\eqref{symm-results}~and~\eqref{non-symm-results},
together with the renormalization constant in \eqref{az-value}, one gets 
for the ratios in Eq.~\eqref{ratios-su3} the values:

\subsubsection*{Vector mesons:}
\eq
\frac{2\alpha_V}{\lambda_V}=0.778\pm 0.014,
\en
\subsubsection*{Baryon octet:}

\eq
\frac{3A_1}{\lambda_B}=0.635\pm0.080\, ,\quad\quad
\frac{3A_2}{\lambda_B}=0.144\pm0.023\, ,\quad\quad
\en
\subsubsection*{Baryon decuplet:}

\eq
\frac{3A}{\lambda_D}=0.590\pm 0.202\, .
\en

These results confirm a behavior for vectors and octet and decuplet baryons close to the quark model prediction,
with deviations compatible with a $1/N_c\simeq 30\%$ correction, as expected from our analysis in Section~\ref{sec:verification}.
In particular, the results for vector mesons are far away from a tetraquark prediction, 
confirming their ordinary $\bar qq$ nature, as it has been known for long from vector meson dominance models~\cite{Gasser:1983yg,Ecker:1988te} 
or the $1/N_c$ expansion~\cite{Pelaez:2003dy,RuizdeElvira:2010cs,Guo:2012ym,Cohen:2014vta,Ledwig:2014cla}.
These results are hence a check of consistency of our method, opening the way to its application to non-ordinary meson candidates,
as for instance the light scalar mesons. 
Nevertheless, this requires a generalization of the Feynman-Hellman theorem for resonances. That is what we will address in the next Section.

\section{Unstable particles}
\label{sec:FH}

In this section, we present a derivation of the analogue of the Feynman-Hellmann theorem
for  resonances in  quantum field theory. To this end, we first consider stable (w.r.t. the
strong interactions)
particles again (say, the Goldstone bosons) and give a derivation in the language of
the Green functions, which does not refer to the eigenstates and eigenvalues of the
QCD Hamiltonian at all.

Let $P^a(x)=\bar q(x)\,i\gamma_5\frac{1}{2}\,\lambda^a q(x)$ 
be a composite field that describes the Goldstone bosons.
Here, $\lambda^a,~a=1,\ldots 8$, are the Gell-Mann flavor matrices. The two-point
function of these fields
\eq\label{eq:D}
D^{ab}(p^2)=i\int d^4xe^{ipx}\langle 0|TP^a(x)P^b(0)|0\rangle
\en
contains a single pole at the physical Goldstone boson mass
\eq\label{eq:pole}
D^{ab}(p^2)\to\frac{\delta^{ab}Z_a}{M_a^2-p^2}+\cdots\, ,\quad\quad
a=1,\cdots 8\, .
\en
Here, the ellipses stand for the terms which are regular at $p^2\to M_a^2$.
For simplicity, we do not consider the case $a=8$ any further, as  $\eta-\eta'$ mixing 
would have to be taken into account.  
The constant $Z_a$ can be expressed via the matrix element of $P^a$ between 
the vacuum and the one-particle state:
\eq
Z_a^{1/2}=\langle 0|P^a(0)|P,a\rangle\, .
\en
Next, we shall use the fact that, in perturbation theory, one can  shuffle the quark 
mass term between the free Lagrangian and the interaction part. We shall use 
this freedom to put {\em all} of the mass term into the interaction. In this case,
the two-point function is given by
\eq\label{eq:perturbation}
\langle 0|TP^a(x)P^b(0)|0\rangle&=&U_0^{-1}
\biggl\langle \Omega_0\biggl|TP_0^a(x)P_0^b(0)
\exp\biggl(i\int d^4y{\cal L}_{\rm int}(y)\biggr)\biggr|\Omega_0\biggr\rangle\, ,
\nonumber\\[2mm]
U_0&=&\biggl\langle \Omega_0\biggl|T\exp\biggl(i\int d^4y{\cal L}_{\rm int}(y)\biggr|\Omega_0\biggr\rangle\, ,
\en
where the interaction part of the QCD Lagrangian is split into the quark mass dependent 
and quark mass independent parts according to
\eq
{\cal L}_{\rm int}(x)={\cal L}_0(x)-\sum_q Z_F^{-1}Z_m m_q^r\bar q_0(x)q_0(x)\, .
\en
Here, $q_0(x)$ is the free {\em massless} quark field, 
$P^a_0(x)=\bar q_0(x)\,i\gamma_5\frac{1}{2}\,\lambda^a q_0(x)$, and
$|\Omega_0\rangle$ denotes the vacuum in the theory with {\em massless} quarks.
The quark mass
independent part of the Lagrangian is not shown explicitly.
Further, $Z_F,Z_m$ denote the quark wave function and quark mass renormalization 
constants, respectively, and $m_q^r$ are the renormalized quark masses. For simplicity,
the minimal subtraction scheme is implied, where $Z_F,Z_m$ do not depend on the
quark flavor and masses.

Let us now ask the question: how do the Goldstone boson masses depend on the
renormalized quark masses $m_q^r(\mu)$, with the scale $\mu$ and the
renormalized coupling constant $g^r(\mu)$ fixed? In order to answer this question, we
shall first differentiate Eq.~(\ref{eq:pole}) with respect to the quark mass $m_q^r$:
\eq\label{eq:diff1}
\frac{\partial D^{ab}(p^2)}{\partial m_q^r}
\to-\frac{Z_a\delta^{ab}}{(M_a^2-p^2)^2}\,\frac{\partial M_a^2}{\partial m_q^r}
+\mbox{single pole}+\mbox{regular}\, .
\en
On the other hand, differentiating Eq.~(\ref{eq:perturbation}) with respect to $m_q^r$,
one obtains
\eq
\frac{\partial D^{ab}(p^2)}{\partial m_q^r}&=&\int d^4xe^{ipx}
\biggl\{U_0^{-1}\biggl\langle \Omega_0\biggl|TP_0^a(x)P_0^b(0)\exp\biggl(i\int d^4y{\cal L}_{\rm int}(y)\biggr)
\nonumber\\[2mm]
&\times&\int d^4z\, Z_F^{-1}Z_m\bar q_0(z)q_0(z)
\biggr|\Omega_0\biggr\rangle
\nonumber\\[2mm]
&-&U_0^{-2}\frac{\partial U_0}{\partial m_q^r}
\biggl\langle \Omega_0\biggl|TP_0^a(x)P_0^b(0)\exp\biggl(i\int d^4y{\cal L}_{\rm int}(y)\biggr)\biggr|\Omega_0\biggr\rangle\biggr\}
\nonumber\\[2mm]
&=&\int d^4x d^4z e^{ip(x-z)}
\biggl\{U_0^{-1}\biggl\langle \Omega_0\biggl|TP_0^a(x)P_0^b(z) 
\exp\biggl(i\int d^4y{\cal L}_{\rm int}(y)\biggr)
\nonumber\\[2mm]
&\times&Z_F^{-1}Z_mS_0(0)
\biggr|\Omega_0\biggr\rangle
\nonumber\\[2mm]
&-&U_0^{-1}\biggl\langle \Omega_0\biggl|TP_0^a(x)P_0^b(z) 
\exp\biggl(i\int d^4y{\cal L}_{\rm int}(y)\biggr)\biggr|\Omega_0\biggr\rangle
\nonumber\\[2mm]
&\times&U_0^{-1}\biggl\langle \Omega_0\biggl|TZ_F^{-1}Z_mS_0(0)
\exp\biggl(i\int d^4y{\cal L}_{\rm int}(y)\biggr)\biggr|\Omega_0\biggr\rangle\biggr\}\, .
\en
Here, $S_0(x)=\bar q_0(x)q_0(x)$ is the {\em (unrenormalized)} scalar density. 
It is now seen that the above equation can be written in the form
\eq
 \frac{\partial D^{ab}(p^2)}{\partial m_q^r}
=\Gamma^{ab}(p,p)\, ,
\en
where
\eq
\Gamma^{ab}(p,q)&=&\int d^4x\, d^4z\, e^{ipx-iqz}
\langle 0|TP^a(x)P^b(z)S^r(0)|0\rangle_{\sf conn}\, ,
\nonumber\\[2mm]
\langle 0|TP^a(x)P^b(z)S^r(0)|0\rangle_{\sf conn}
&=&\langle 0|TP^a(x)P^b(z)S^r(0)|0\rangle
\nonumber\\[2mm]
&-&\langle 0|TP^a(x)P^b(z)|0\rangle\langle 0|TS^r(0)|0\rangle\, ,
\en
and $S^r(x)=Z_F^{-1}Z_mS_0(x)$ denotes the renormalized scalar density.

Further, inserting a complete set of states in the above equation, it is straightforward 
to see that the quantity $\Gamma^{ab}(p,q)$ contains a double pole in the variables
$p^2,q^2$
\eq\label{eq:diff2}
\Gamma^{ab}(p,q)\to -\frac{\delta^{ab}\langle 0|P^a(0)|P,a\rangle
\langle P,a|S^r(0)|P,b\rangle\langle P,b| P^b(0)|0\rangle}{(M_a^2-p^2)(M_a^2-q^2)}
+\cdots\, ,
\en
where the ellipses denote the less singular terms. Comparing the coefficients
in front of the double pole in Eqs.~(\ref{eq:diff1}) and (\ref{eq:diff2}), we
finally arrive at the Feynman-Hellmann theorem
\eq
\frac{\partial M_a^2}{\partial m_q^r}=\langle P,a|S^r(0)|P,a\rangle\, .
\en
As already mentioned, the advantage of the this derivation as compared to the 
standard one is that it does not refer to the (real) eigenvalues of the Hamiltonian
$E_n(\lambda)$ from the beginning, dealing instead with the Green functions in QCD.
Hence, this derivation 
can be directly generalized to the resonances\footnote{In the above derivation, 
one might feel slightly uncomfortable
with the trick that introduces massless quarks at the intermediate stage of the proof,
albeit no reference to these is left in the final expression. This (superficial) problem 
can be, however, easily avoided. Instead of putting whole quark mass term into the perturbation, one could, e.g., introduce the formal parameters
$s_q=m_q-m_q^{phys}$. The derivative in the Feynman-Hellmann theorem is taken with respect to the parameters $s_q$ at $s_q=0$, and one arrives at the same result at the 
end. In order to keep the notations as simple as possible, we refrained from introducing
the additional parameters $s_q$.}. In the latter case, the two-point function
is defined by the expression similar to Eq.~(\ref{eq:D}) where, instead of $P^a(x)$,
one may use any operator with the quantum number of a given resonance.
Then, the two-point function has a pole on some unphysical Riemann sheet of the
complex $p^2$-plane and not on the real axis. By the same token, the matrix element
of the operator $S^r(x)$ between the resonance ``states'' is defined similarly
to Eq.~(\ref{eq:diff2}), through the residue of the three-point function at the 
double pole\footnote{These quantities can be extracted from the lattice data, 
see~\cite{Bernard:2012bi,Agadjanov:2014kha,Agadjanov:2016fbd}. As an example of 
calculation of the resonance matrix element in ChPT, we refer to~\cite{Albaladejo:2012te}. 
The quark mass dependence of the $\sigma$-meson pole in the unitarized ChPT
has been addressed in Ref.~\cite{Bruns:2016zer}}.
Consequently, the only difference between the stable states and the resonances
boils down to the question, whether a pole is real or not. This difference is inessential 
for the derivation of the Feynman-Hellmann theorem, which is given above. Consequently,
it still holds, if one interprets $M_a^2$ as a resonance pole position in the complex
plane and not as the energy of an isolated energy level on the lattice. Furthermore,
the large-$N_c$ and SU(3) symmetry arguments apply to the Green functions 
irrespective of the fact,
whether they have a real or a complex pole. Consequently, the physical meaning of
the parameters $\gamma,\gamma',\beta,\beta'$ remains the same, albeit they
become complex for resonances. Namely, if these happen to be close to the (real)
quark model values for a given multiplet, then this multiplet has a little admixture
of exotica and vice versa.

One may also wonder, whether the Gell-Mann--Okubo formula, which was 
extensively used above, is applicable in the case of the resonances. The answer to this
question can be found along the similar pattern. The quark mass term in the Lagrangian
can be rewritten as (to ease notations, the renormalization constants are suppressed)
\eq
{\cal L}_m=\hat m(\bar uu+\bar dd)+m_s\bar ss=\bar m(\bar uu +\bar dd +\bar ss)
+\lambda \frac{m_1}{3}\,(\bar uu +\bar dd -2\bar ss)\, ,
\en
where $\lambda=1$ in the real world. Suppose now that, for any $\lambda$,
one has a multiplet of poles $M_a^2$ (real or complex) in the two-point
function of the operators with appropriate quantum numbers. 
One could now differentiate with respect to the parameter $\lambda$ and get
\eq
\frac{\partial M_a^2}{\partial\lambda}=\langle P,a|O_8^r(0)|P,a\rangle\, ,
\en
where $O_8^r$ denotes the renormalized operator proportional to
$\bar uu +\bar dd -2\bar ss$. In case of the resonances, the matrix element in the right-hand side is understood, 
as the residue of the pertinent three-point function at the complex pole.

The above relation is written for any $\lambda$. For $\lambda=1$, we are back to
the real world. One may first consider it for $\lambda\to 0$, where it yields 
the first-order correction to the SU(3)-symmetric limit.
The Lagrangian is explicitly SU(3) symmetric, as $\lambda\to 0$, and both the interpolating particle fields and the operator $O_8^r$ transform as irreducible tensor
operators of SU(3). The group-theoretical analysis applies directly to the three-point function, and the Gell-Mann--Okubo formula holds -- even for the resonances.

Continuation to $\lambda=1$, i.e., back to the real world, is a subtle issue. The 
(approximate) validity of the Gell-Mann--Okubo formula is, in fact, equivalent to the 
statement that the linear term in $\lambda$ describes the spectrum well up to
$\lambda=1$. In case of stable particles, there exists no internal 
contradiction in assuming
this. Consider, however, the situation, when all particles in the multiplet are stable
at the SU(3)-symmetry point. Increasing $\lambda$ introduces the mass splitting,
and some of the particles become unstable at $\lambda=\lambda_{crit}<1$. Assuming
the analyticity in $\lambda$ then leads to the controversy since, as it is well known,
the real and imaginary parts of the pole position have cusps at threshold. Consequently, 
the assumption that the Gell-Mann--Okubo formula approximately holds for the 
resonance 
masses as well implies that the cusp effects are small and hence, this approximation
should work better for the resonances with higher spin. 

Finally, a few words about testing the exotic content of the resonances with the use of the
lattice simulations. As it is well known, the resonances do not correspond to the individual
energy levels of the lattice QCD spectrum. In order to extract the position of the 
resonance pole, one has to first determine the phase shift at a given energy by use
of the L\"uscher equation and, at the next stage, find the pole position through the
extrapolation into the complex energy plane.  Recent years have seen some progress
in this direction, see, e.g., Refs.~\cite{Briceno:2016mjc,Wilson:2015dqa}. More work
is, however, necessary to perform a full-fledged investigation  of  exotic resonances
on the lattice.

\section{Conclusions}
\label{sec:concl}

\begin{itemize}

\item[i)]
We propose a criterion which allows one to judge, whether the hadrons in a given
multiplet are predominately quark-model states or exotic states. The quantities
$\gamma,\gamma',\beta,\beta'$ for different multiplets are observable quantities, 
expressed through the $\sigma$-terms.
In the quark model, these quantities are exactly given by the group-theoretical factors.
Should it turn out that the values of these quantities for some multiplet 
in the real world significantly differ
from the quark model values, one would interpret this as a signature of the exotic
character of a multiplet in question.    

\item[ii)]
The above criterion has been verified, using Chiral Perturbation Theory and large-$N_c$
arguments. It has been shown that the quark model values for  
$\gamma,\gamma',\beta,\beta'$ emerge in QCD at the leading order in $1/N_c$, both in
the meson and in the baryon sector.

\item[iii)]
Using the hadron mass values, measured at different input values of the quark masses
in lattice QCD simulations, we have verified our criterion in case of the 
pseudoscalar and vector meson octets, as well as the low-lying baryon octet 
and decuplet.
As expected, the fit to the lattice data gives results close to the quark model 
predictions. In some parameters, the difference of order of $1/N_c^{}\simeq 30\%$
is observed. It will be extremely important to apply the same criterion to the lightest 
scalar
meson octet, which is the most obvious candidate for the low-lying exotic multiplet.

\item[iv)]
The Feynman-Hellmann theorem, which has been used above to calculate the
quantities $\gamma,\gamma',\beta,\beta'$, has been generalized for the resonance
states. A field-theoretical proof is provided.
 The criterion for the exotic multiplets does not change
its form. Such a generalization is necessary, because all candidates
for QCD exotica are resonances and not stable particles.  

\end{itemize}

\subsection*{Acknowledgments}

The authors thank J.~M.~Alarcon, M. Frink, J. Gasser, H. Leutwyler,
F.~J.~Lla\-nes-Estrada, J. Nebreda, 
A. Manohar and J.~R.~Pelaez for useful discussions.
GS thanks his colleagues of the QCDSF Collaboration for sharing the lattice results.
 We acknowledge the support from the DFG (CRC 110 
``Symmetries and the Emergence of Structure in QCD'').
This research is supported in part by Volkswagenstiftung under contract no. 86260,
by the Chinese Academy of Sciences (CAS) President's International Fellowship 
Initiative (PIFI) (Grant No.\! 2017VMA0025), the Swiss National Science Foundation (SNF), 
and by Shota Rustaveli National Science Foundation (SRNSF), grant no. DI-2016-26.

\end{document}